\documentclass[12pt]{iopart}

\usepackage{graphicx}
\usepackage{epstopdf}

\begin{document}

\title{Fully Developed Turbulence in the view of Horizontal Visibility Graphs}

\author{Pouya Manshour$^1$, M. Reza Rahimi Tabar$^{2,3}$ and Joachim Peinke$^3$}

\address{
$^1$Department of Physics, Persian Gulf University, 75169 Bushehr, Iran\\
$^2$Department of Physics, Sharif University of Technology, 11155-9161 Tehran, Iran\\
$^3$Institute of Physics, University of Oldenburg, 26129 Oldenburg, Germany
}
\ead{manshour@pgu.ac.ir}ý

\begin{abstract}
We employ the horizontal visibility algorithm to map the velocity and acceleration time series in turbulent flows with different Reynolds numbers, onto complex networks. The universal nature of velocity fluctuations in high Reynolds turbulent Helium flow is found to be inherited in the corresponding network topology. The degree distributions of the acceleration series are shown to have stretched exponential forms with the Reynolds number dependent fitting parameter. Furthermore, for acceleration time series, we find a transitional behavior in terms of the Reynolds number in all network features which is in agreement with recent empirical studies.
\end{abstract}

\maketitle

\section{Introduction}
\label{intro}
Turbulence is of key interest from the point of view of both non-linear dynamical systems and irreversible statistical mechanics. As the velocity of a fluid exceeds some critical value, the stationarity and regularity of the flow break off, and small velocity disturbances are no longer damped by the laminar flow, but grow by extracting kinetic energy from the mean flow. In this situation, which is known as turbulent flow regime, fluid particles follow complex trajectories with stochastic dynamics, and their velocity gradients are much larger than in the laminar case \cite{Frisch1995}. The transition to the turbulence is related to the supremacy of the inertial forces over the viscous forces. Reynolds proposed to quantify the competition among these two mechanisms by a control parameter, called Reynolds number \cite{Reynolds1883}, defined as $Re\equiv UL/\nu$, where $U$, $L$, and $\nu$ are the mean velocity, the characteristic length scale of the flow, and the kinetic viscosity, respectively. Thus, the turbulent regime corresponds to high $Re$ numbers.

A turbulent flow is characterized by a hierarchy of scales, i.e., a flux of energy is injected into the fluid motion at large scales and is continuously transported towards smaller scales, $r$, and consequently is dissipated by the molecular viscosity at the smallest scale, called \emph{Kolmogorov scale} \cite{Frisch1995}. This is well known as the \emph{energy cascade} phenomena in turbulence. These two scales at the extremes of the cascade process can defer by several orders of magnitude at high Reynolds number.

A common way to study the statistics of turbulent velocity fields is by means of the longitudinal velocity increments at time $t$ and location $\textbf{x}$, $\delta u(r,t)=\textbf{e}\cdot [\textbf{u}(\textbf{x}+\textbf{e}r,t)-\textbf{u}(\textbf{x},t)]$, where $\textbf{e}$ denotes an arbitrary direction unit vector. At large scales, $r\approx L$, fluid motions are statistically independent and the probability distribution function (PDF) of the velocity increments is found to be nearly Gaussian. At scales $r<L$, turbulent motions become \emph{intermittent}, in which rather quiescent periods in the velocity signal are interrupted irregularly by strong correlated bursts \cite{Batchelor1949}. As a consequence, the PDF of $\delta u(r)$ develops long tails and becomes strongly non-Gaussian \cite{Sreenivasan1997,Friedrich2011}. A generally accepted way of describing the intermittency is by means of the \emph{flatness}, defined by $F\equiv\langle\delta u(r)^4\rangle/\langle\delta u(r)^2\rangle^2$, where $F=3$ is for a Gaussian and $F>3$ is for a non-Gaussian PDF. Empirical measurements in various geometries \cite{Sreenivasan1997} indicate a power-law increase, $F\sim {R_{\lambda}}^{0.35}$, where $R_\lambda$ denotes the Taylor microscale Reynolds number, defined as $R_{\lambda}=u'\lambda/\nu$, where $u'$ is the velocity fluctuation, $\lambda$ is the Taylor scale, and $\nu$ is the viscosity. This power-law behavior is irrespective of the flow, and also shows the increase in the level of intermittency, which is also consistent with phenomenological intermittency models \cite{Sreenivasan1997,Meneveau1991,Grossmann1992}.

Acceleration, $a(t)=du(t)/dt$, in fluid flow where causes irregular fluctuating motions of fluid particles, is also another essential quantity in turbulence studies \cite{vedula1999,voth1998,porta2001,voth2002,Biferale2004,reynolds2004}, and has a crucial role in cloud formation and atmospheric transport, processes in stirred chemical reactors and combustion systems and in the industrial production of nanoparticles \cite{Vaillancourt2000,Weil1992,pope1994,pratsinis1996, Farzaneh}. For example, recent experimental studies on the fluid particle accelerations in fully developed turbulence \cite{voth1998,porta2001,voth2002} have shown that the fluid acceleration occurs in a very intermittent fashion, and the probability density functions of the acceleration at different Reynolds numbers exhibit stretched exponential tails. The pressure gradient fluctuations, the coherent motions of fluids, or the rotational motion of the vortices are some suggested sources of the anomalous scaling in the acceleration statistics \cite{vedula1999,voth2002,Sawford2003}.

It is also shown that the multifractal formalism can predict the fat-tail nature of the acceleration PDFs \cite{Biferale2004}. As the Reynolds number increases, the role of the intermittency becomes more and more important. Phenomenological models predict a power-law increase with $R_\lambda$ \cite{vedula1999,reynolds2004,Sawford2003} for the acceleration variance, $\sigma_a\sim R_\lambda^{0.14}$. It also has been shown \cite{Hill1995} that the acceleration variance, $\sigma_a$, is closely related to the flatness, $F$. However, such power-law behaviors in the flatness and variance of acceleration have been challenged by some experimental studies \cite{voth1998,porta2001,Tabeling1996,Tabeling2002}. Their detailed measurements show that these quantities first increase with Reynolds number up to $R_{\lambda}\simeq 700$ and then cease to increase further. The particle size effect \cite{voth1998} or the characteristics of a second order phase transition based on worm vortex breakdown \cite{Tabeling2002} are some proposed origins of such a transitional behavior.

In spite of the possibility of finding some particular solutions of Navier-Stokes equation governing fluid motion, all such solutions are unstable to finite perturbations at high Reynolds number. In other words, random velocity fluctuations in a wide range of different length and time scales and the intermittent nature of the fully developed turbulent flows pose profound problems in the numerical and theoretical analysis of the turbulence dynamics for experimentalists and theoreticians, which makes the search for new approaches inevitable.

Recently, the complex network theory has been proved itself as a main tool to study the complex systems \cite{Barabasi2002}, and has helped understand many important processes in physics, biology, neuroscience, communications, epidemiology among others \cite{Barabasi2002,Barrat2008,Satorras2007,Murray1993,Dezso2002,Montakhab2012,Manshour2014}. By using the useful concepts from graph
theory as well as statistical physics, many advances have been made to find and predict various interesting features of complex systems. Consequently, in recent
years, many efforts have been done to map a time series onto a graph, and then by studying the properties of the graph, one may extract some features of the
corresponding time series from the resulting graph \cite{Zhang2006,Lacasa2008,Lacasa2009pre,Shirazi2009,Telesca2012,marwan2009}, for a review see \cite{Donner2011}.

 In this respect, Lacasa \textit{et al.} introduced an algorithm \cite{Lacasa2008}, called visibility graph (VG), which maps a time series into a graph based on the
ability of the data points to see each other and it is defined as follows: Let $x_i$ be a univariate time series of $N$ data ($i=1,2, ... N$).
The algorithm assigns each datum of the series to a node in the VG. Accordingly, a series of size $N$ maps to a graph with $N$ nodes. Two nodes $i$ and $j$ in
the graph are connected if one can draw a (straight) line in the time series joining $x_i$ and $x_j$ that does not intersect any intermediate data height. That is two arbitrary data values $(t_i, x_i)$ and $(t_j, x_j)$ will have visibility, and consequently will become two connected nodes of the associated graph,
if any other data $(t_q, x_q)$ placed between them satisfies:
\begin{equation}
x_q<x_j + (x_i-x_j)\frac{t_j-t_q}{t_j-t_i}.
\label{vg}
\end{equation}

Note that the visibility graph is always connected by definition and also is invariant under affine transformations, due to the mapping method.
An alternative (and much simpler) algorithm is the horizontal visibility graph (HVG) \cite{Lacasa2009pre}, in which a connection can be established between two data
points $i$ and $j$, if one can draw a \textit{horizontal} line in the time series joining them that does not intersect any intermediate data height, $x_q$,
by the following geometrical criterion:
\begin{equation}
x_i,x_j>x_q \mbox{  for all $q$ such that $t_i < t_q < t_j$}.
\label{hvg}
\end{equation}

Because of the simplicity of the HVG, some features of the graph can be analytically calculated. It has been shown in \cite{Lacasa2009pre} that for an uncorrelated time series, the corresponding HVG is small-world network with mean degree $\left\langle k\right\rangle=4$ and also its degree distribution, $p_k$ is as follows:
\begin{equation}
p_k \sim e^{-\lambda_c k}
\label{exp}
\end{equation}
with $\lambda_c=\ln(3/2)$ and these results are universal, i.e. \emph{independent} of the probability distribution from which the series was generated.
On the other hand, ordered (periodic) series convert into regular graphs: thus order and disorder structure in the time series seem to be inherited in the
topology of the visibility graph. It was also shown that for a fractal time series, the distributions, associated to the resulting visibility graphs, follow a power law $p_k \sim k^{-\gamma}$ \cite{Lacasa2009epl}, such that the Hurst exponent $H$ of the series is linearly related to $\gamma$. The visibility algorithm has been also proposed to study the distinction between chaotic and stochastic time series \cite{Lacasa2010}, which has been recently proven not to hold in general \cite{Ravetti2014}.

In this article, we apply the HVG algorithm to map the velocity, and the acceleration time series in turbulent flows with various Reynolds numbers,
onto complex networks. Then, we provide a distinct perspective on the temporal correlation information in the time series by calculating different
topological properties of the HVG. At first, we demonstrate that the universality of velocity statistics at high Reynolds numbers is also present in the corresponding HVG degree distributions. Then we focus our attention on the acceleration time series, and will show that the stretched exponential functions can well be fitted to the degree distributions and the corresponding fitting parameter has a Reynolds number dependency. By taking into account other topological measures, such as the standard deviation of the degrees, the degree assortativity, and Spearman's correlation coefficient between the node's degree and its corresponding value in the original time series, we will demonstrate that these topological measures allow us to discriminate between the turbulence time series with different Reynolds numbers. Furthermore, at high Reynolds numbers we observe a transitional behavior for all of the above mentioned topological properties for the acceleration time series, which has been observed in the recent empirical studies \cite{voth1998,porta2001,Tabeling1996,Tabeling2002}.

\begin{figure}[t]
\begin{center}
\includegraphics[scale=0.6]{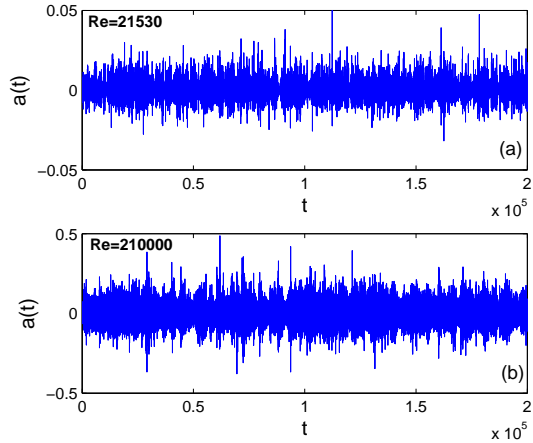}
\caption{The two acceleration time series corresponding to the different Reynolds numbers of (a) $21530$ and (b) $210000$.}
\label{orig_series}
\end{center}
\end{figure}

\begin{figure}[t]
\begin{center}
\includegraphics[scale=0.5]{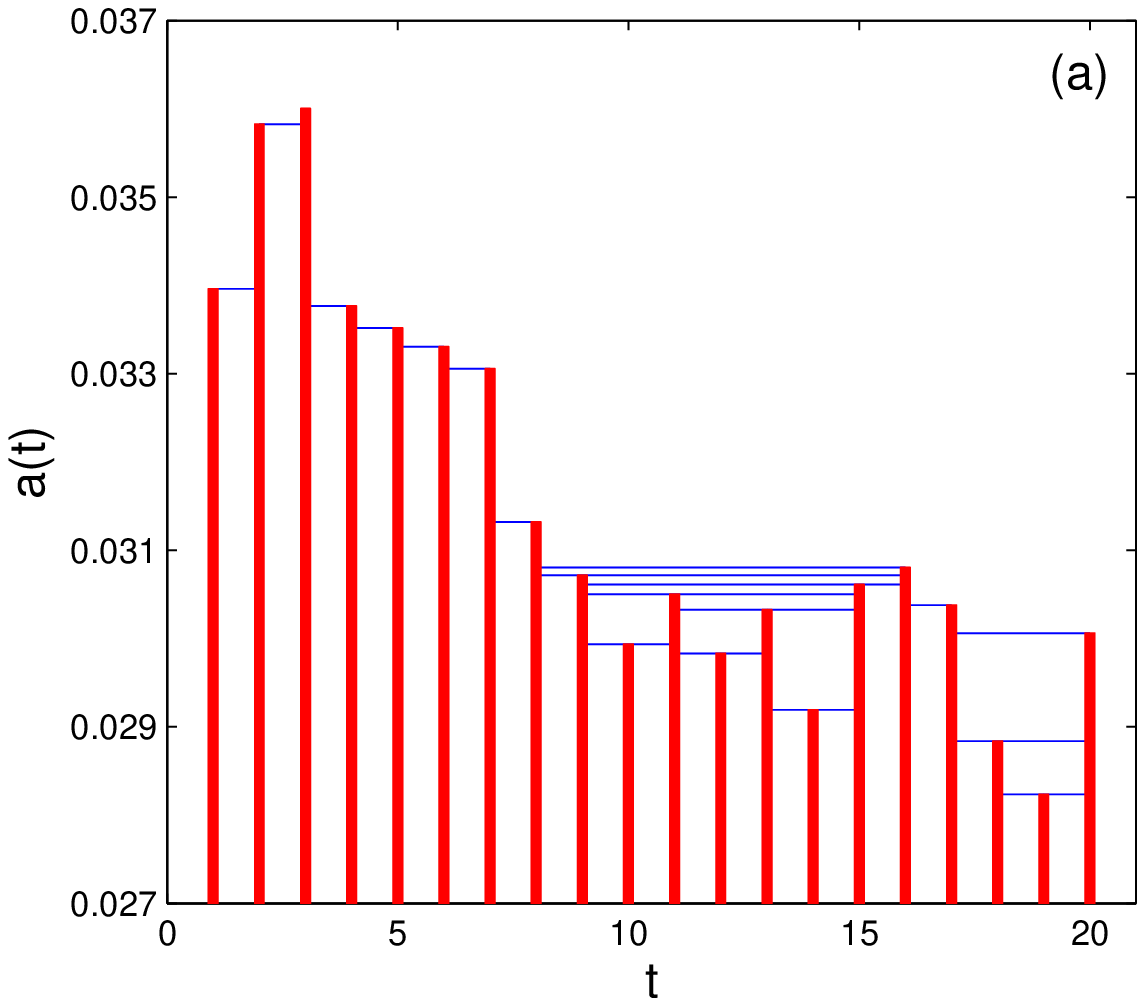}
\includegraphics[scale=0.5]{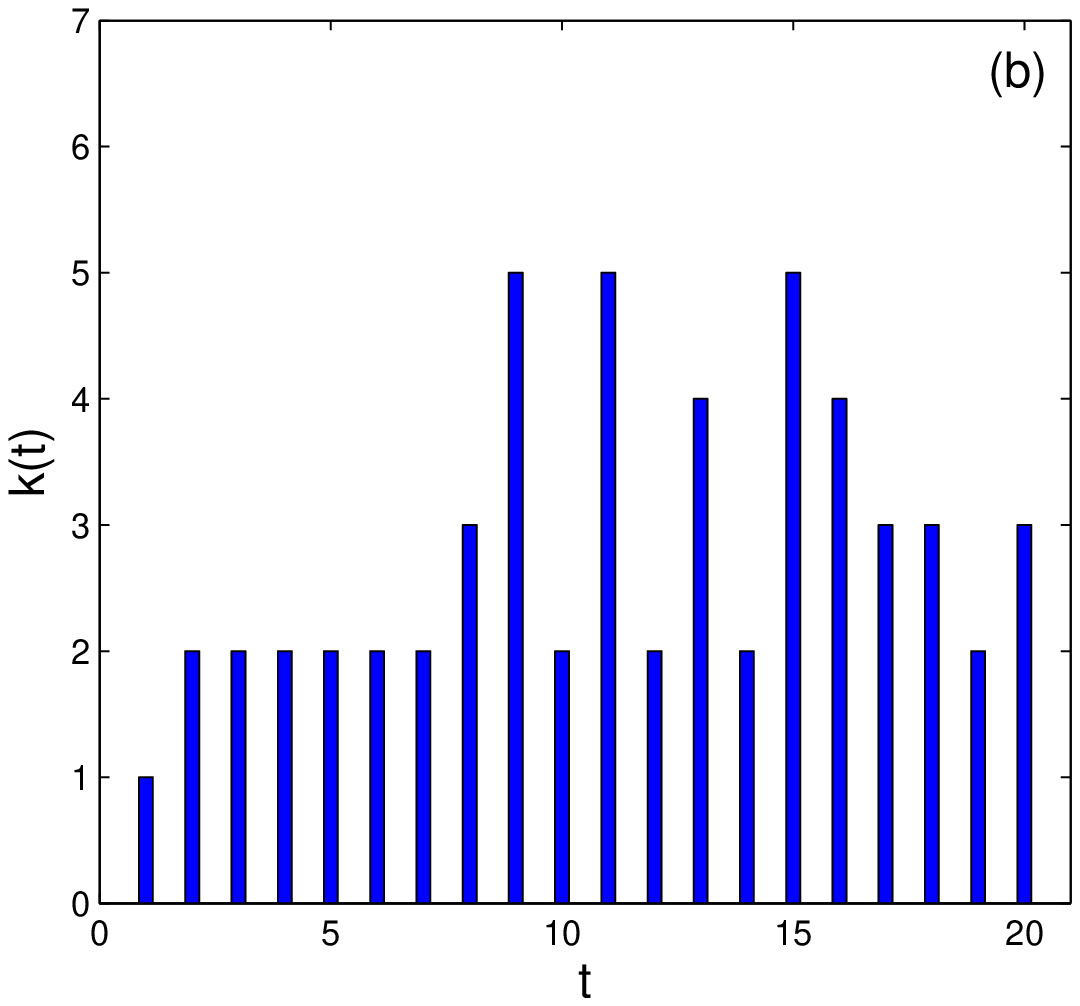}
\caption{(a) the first $20$ data points of the acceleration time series for Reynolds number of $21530$ and its corresponding HVG connectivity links. (b) the corresponding HVG degree sequence for such data points. It is obvious that a global high-value data point does not necessarily correspond to a high degree node (e.g. at $t=3$).}
\label{x_t}
\end{center}
\end{figure}

\section{Results and discussions}

\subsection{Data description}

We use the experimental results \cite{Renner2001} of hot-wire measurements in the central region of an air into air round free jet low temperature helium turbulence with Reynolds numbers $21530$, $48190$, $115000$, $210000$, and $757000$, were obtained by a nozzle with an opening diameter of $D=8$ mm. The closed experimental chamber is $2.5$ m high with a cross-section of $1$ $m^2$, which guarantees that a turbulent jet does not interact with the walls up to a distance of more than $150$ nozzle diameters. The time-resolved measurements of the local velocity in the direction of the mean flow were performed by means of hot-wire anemometry, using a single wire probe with a spatial resolution of $1.25$ mm and a time resolution of $8$ kHz \cite{Renner2001}. We apply the HVG algorithm to map the velocity and the acceleration time series of size $N \sim 10^6$ onto a network.

\subsection{Degree sequence}
For example, in Fig.~\ref{orig_series}, we plotted two acceleration time series with different Reynolds numbers of $21530$ and $210000$. As we described before, there is an one-to-one correspondence between the time order of data points of the time series and the node's indices in the corresponding graph. Fig.~\ref{x_t} illustrates the HVG mapping formalism, and shows the first $20$ acceleration data points, $a(t)$, belongs to Reynolds number $21530$ and its corresponding HVG degree sequence, $k(t)$. As it can be seen, two data points satisfying the horizontal visibility condition are connected by a line indicating the presence of an edge in the HVG. Note that a high degree node in the HVG does not necessarily correspond to a \emph{globally} high-value data point in the original time series. In Fig.~\ref{x_t}, we can see that the local maximum data point $a(t=11)$ has high degree in spite of its \emph{globally} low value in the original time series. It is noteworthy to mention here that the actual degrees are more likely to have larger values for time series with larger sizes, since HVGs associated to short time series suffer from severe edge effects which lead to a systematic downward bias of the degrees for nodes close to both ends of the series \cite{Donner2012}.
\subsection{Degree distribution}

\begin{figure}[t]
\begin{center}
\includegraphics[scale=0.5]{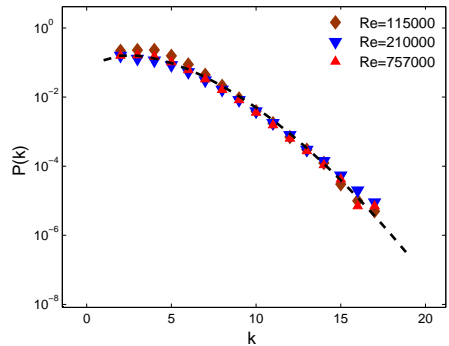}
\caption{The HVG degree distributions of the velocity time series with three different high Reynolds numbers $115000$, $210000$, and $757000$.
The dashed line shows the stretched exponential function of Eq.~\ref{PDF}. The universality of velocity statistics in high Reynolds numbers is inherited in the HVG degree distributions.}
\label{degree_dist_vel}
\end{center}
\end{figure}

In Fig.~\ref{degree_dist_vel}, we plotted the HVG degree distributions for the velocity time series, $u(t)$, for  three (high) Reynolds numbers of $115000$, $210000$, and $757000$. Clearly, a universal behavior is present in this plot. We propose to model such degree distributions with the stretched exponential functions (SEF) \cite{Stretched}:
\begin{equation}
P(k)=\gamma k^{\gamma-1}/k_0^\gamma exp(-(k/k_0)^\gamma)
\label{PDF}
\end{equation}
where $\gamma$ and $k_0$ are the fitting parameters. The dashed line in Fig.~\ref{degree_dist_vel} represents a fitted function with parameters $k_0\simeq4$, and $\gamma \simeq 1.75$, obtained from the data regression analysis.

\begin{figure}[t]
\begin{center}
\includegraphics[scale=0.5]{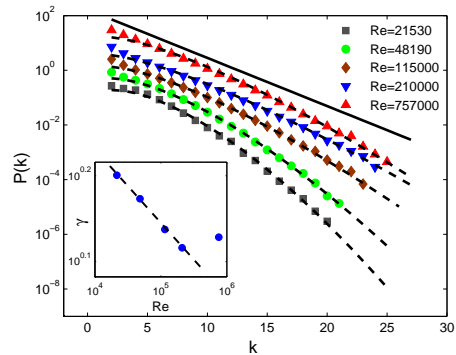}
\caption{The HVG degree distributions of acceleration time series for five Reynolds numbers $21530$, $48190$, $115000$, $210000$, and $757000$.
The dashed lines show the stretched exponential functions, Eq.~\ref{PDF}. A simple exponential function of Eq.~\ref{exp} has been plotted for better comparison (solid line). The curves are shifted in the vertical direction for convenience of presentation. The inset also shows the fitting parameter, $\gamma$, for different values of the Reynolds numbers.}
\label{degree_dist}
\end{center}
\end{figure}

Let us concentrate on the corresponding HVGs of the acceleration time series. Fig.~\ref{degree_dist} demonstrates the degree distributions of the corresponding acceleration time series for five Reynolds numbers of $21530$, $48190$, $115000$, $210000$, and $757000$. The solid line shows the exponential function of Eq.~\ref{exp}, for better comparison. We observe that the probability of finding high degree nodes (the tails) increases with Reynolds number. The dashed lines in Fig.~\ref{degree_dist} represent the fitted stretched exponential functions of Eq.~\ref{PDF} and we find that $k_0\simeq4$, independent of $Re$ and the exponent $\gamma$ is a Reynolds number-dependent parameter (see Table~\ref{table1}). The inset in Fig.~\ref{degree_dist} shows such dependency of the form of $\gamma \sim {Re}^{-\beta}$, in a logarithmic scale, with $\beta \simeq-0.1$. Note that, in very high Reynolds numbers, for instance above $250000$ (which is corresponds to $R_{\lambda}\simeq 650$ \cite{Renner2001}), the $\gamma$ cease to decrease further. Observing such transitional behavior is in agreement with the reported results in the literatures \cite{voth1998,porta2001,Tabeling1996,Tabeling2002}.

A frequently studied property in complex network theory is the degree variance, $\sigma_k^2=\left\langle k^2\right\rangle-\left\langle k\right\rangle^2$, which can be calculated straightforwardly from Eq.~\ref{PDF}. For a nearly constant value of $k_0$, increasing $\gamma$ corresponds to decreasing $\sigma_k$. In Table~\ref{table1}, we represent the calculated values of the standard deviation of the degrees, $\sigma_k$, versus the Reynolds numbers. As it can be seen, the fluctuations in degrees increase with Reynolds number.

\begin{table}
\caption {The calculated values of five topological parameters for five different Reynolds numbers, using HVG algorithm.}
\label{table1}
\begin{center}
\begin{tabular}{|c|c|c|c|c|c|c|c|}
\hline
$Re$     & $\gamma$ & $\sigma_k$ &  $r$     & $S$      & $\rho_{A}$  \\
\hline
\hline
$21530$  & $1.5870$ & $1.9318$   & $0.0245$ & $0.3367$ & $-0.2982$   \\
$48190$  & $1.5010$ & $2.1396$   & $0.0553$ & $0.4557$ & $-0.3586$   \\
$115000$ & $1.3721$ & $2.3760$   & $0.1139$ & $0.6369$ & $-0.3900$   \\
$210000$ & $1.3064$ & $2.4571$   & $0.1507$ & $0.7432$ & $-0.3921$   \\
$757000$ & $1.3440$ & $2.4120$   & $0.1574$ & $0.7060$ & $-0.3612$   \\
\hline
\end{tabular}
\end{center}
\end{table}

\subsection{Linear/nonlinear correlation effects on the degree distribution}
\begin{figure}[t]
\begin{center}
\includegraphics[scale=0.5]{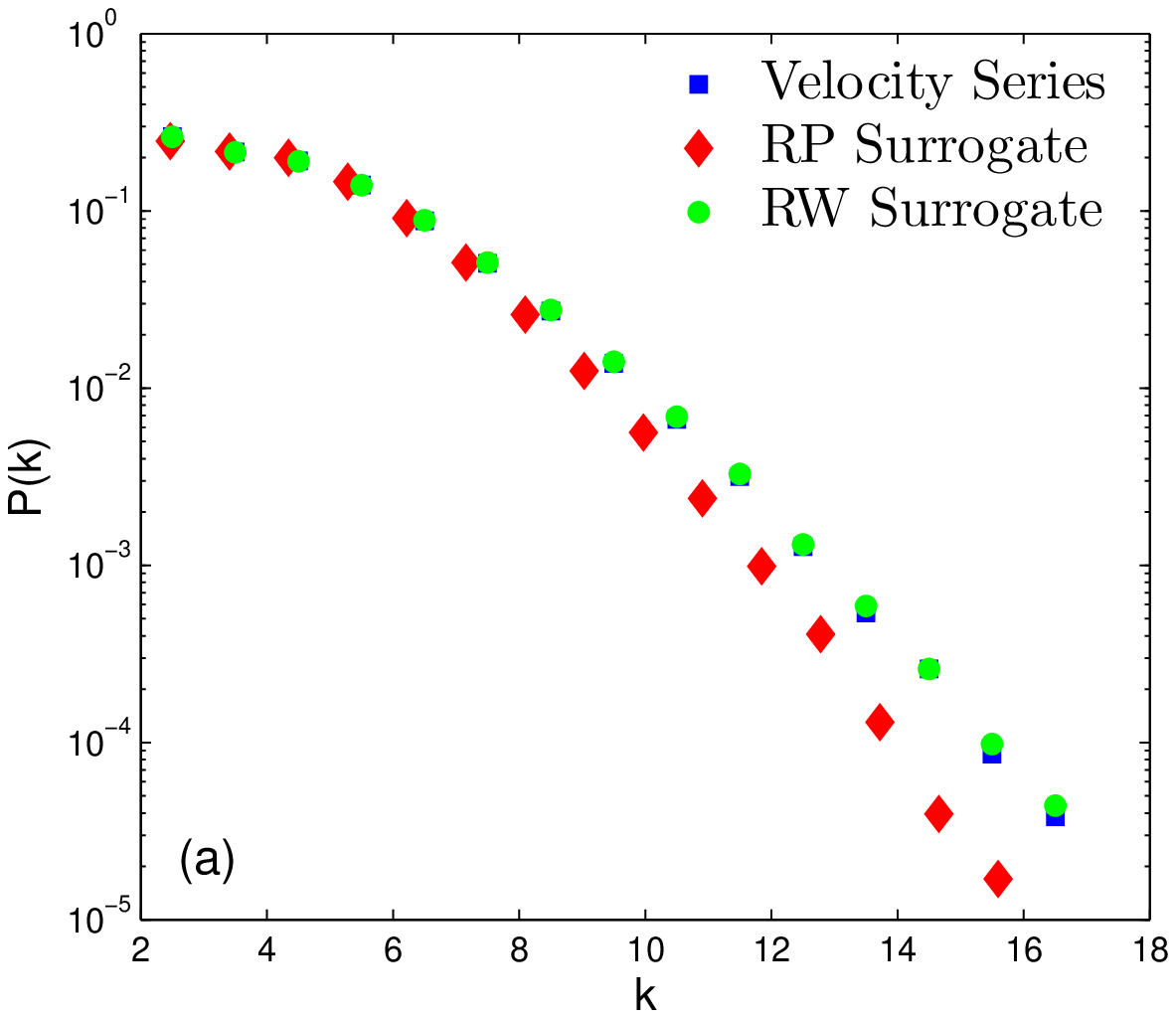}
\includegraphics[scale=0.5]{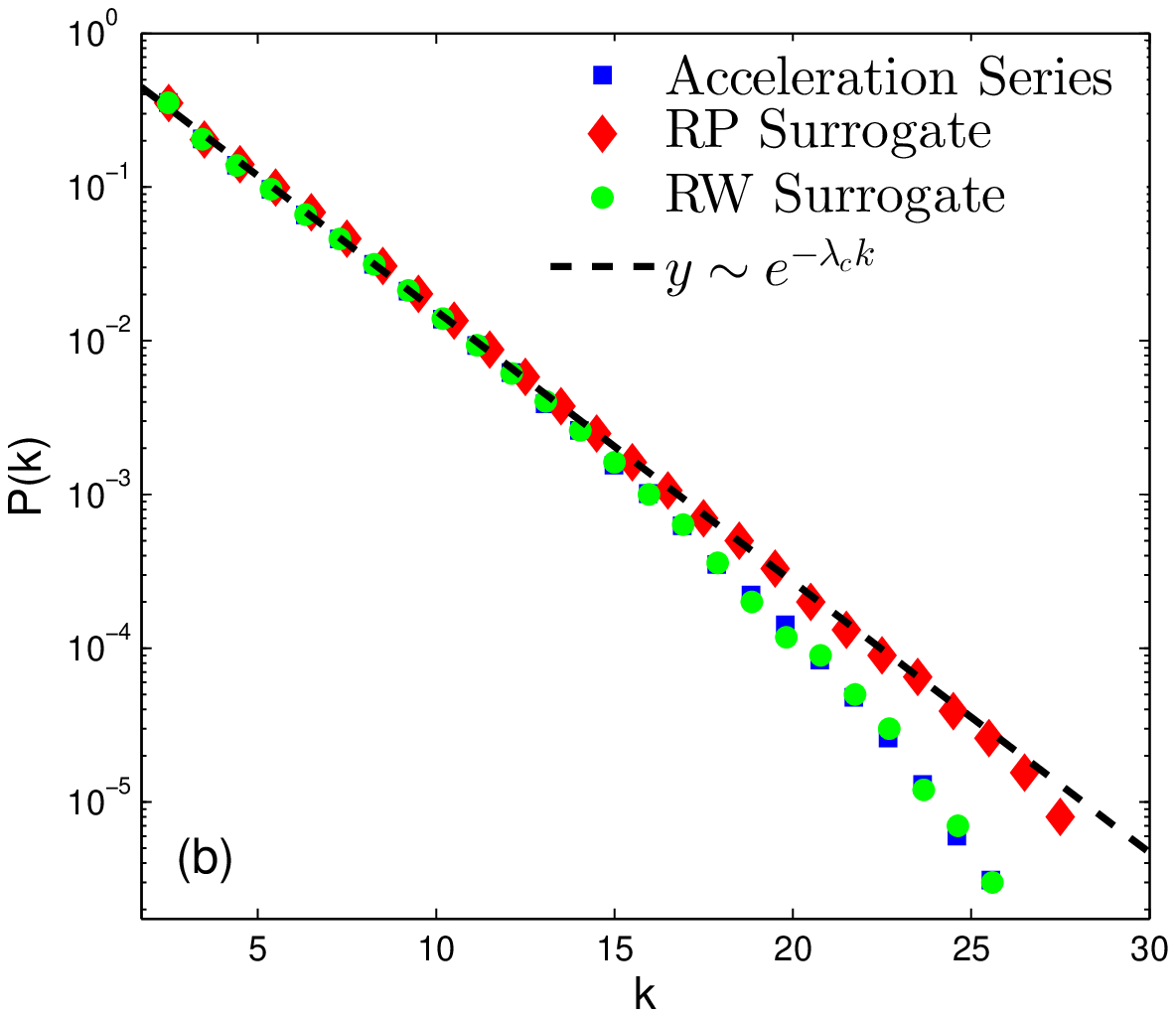}
\caption{(a) The HVG degree distributions of the velocity time series for Reynolds number $210000$ (squares), and its corresponding random phase (diamonds) and rank-wise (circles) surrogate data. (b) As in (a), but for the acceleration time series. The dashed line in (b) shows a simple exponential function of Eq.~\ref{exp}.}
\label{surrog}
\end{center}
\end{figure}

As we mentioned above, the HVG algorithm is independent of the probability distribution of the original time series. Therefore, it is also interesting to see how the network topology depends on the linear and nonlinear correlation features of the original process. To check the effects of such features on the HVG statistics, we use two different methods for surrogating data. The first is to exchange the time series rank-wise (RW) by a Gaussian distributed one \cite{Bogachev2007}, in which the (linear and nonlinear) correlation structure is conserved, but the distributional effect is eliminated. The second one is the phase randomization (RP), where only linear correlations remain in the process, and the nonlinear memory as well as the distributional effects are eliminated \cite{Schreiber1996}. In Fig.~\ref{surrog}, we plotted the HVG degree distributions of velocity (Fig.~\ref{surrog}(a)) and acceleration (Fig.~\ref{surrog}(b)) time series with Reynolds number $210000$, and of their associated rank-wise and phase-randomized surrogate data. We can see that RW surrogation does not affect HVG statistics, as expected. However, for RP surrogation a significant change appears in both time series. For example, since acceleration time series has only nonlinear correlation (with negligible linear correlation), its corresponding degree distribution due to the RP surrogation method becomes the exponential function of Eq.~\ref{exp}. The presence of any differences between the HVG degree distributions associated to the random-phase and rank-wise surrogate data, is an indicator of the existence of nonlinear correlation. In summary, the HVG degree distributions can capture the linear and nonlinear correlation structure in the turbulence time series.

\subsection{Assortativity and Spearman's coefficient}
\label{assortativity}
The assortativity coefficient $r$ is the Pearson's correlation coefficient between the degrees of pairs of linked nodes \cite{Newman2002}.
The excess degree of a node, which is the number of edges leaving the node other than the one we arrived along, is distributed according to
\begin{equation}
q_k={(k+1)p_{k+1}}/{\left\langle k \right\rangle}
\label{excess_degree}
\end{equation}
where $p_k$ is the degree distribution and $\left\langle k \right\rangle =\Sigma_k kp_k$ is the mean degree in the network.
The assortativity coefficient for mixing by vertex degree in an undirected network is
\begin{equation}
r=\frac{\Sigma_{jk}jk(e_{jk}-q_jq_k)}{\sigma^2_q}
\label{assort}
\end{equation}
where $e_{jk}$ is the fraction of edges that connect vertices of degrees $j$ and $k$, and $\sigma_q$ is the standard deviation of the
distribution $q_k$ \cite{Newman2002}. Positive values of $r$ indicate a correlation between nodes of similar degree, while negative values indicate
relationships between nodes of different degree. In general, $r$ lies between $-1$ and $1$. When $r=1$, the network is said to have perfect
assortative mixing patterns, when $r=0$ the network is non-assortative, while at $r=-1$ the network is completely disassortative.

\begin{figure}[t]
\begin{center}
\includegraphics[scale=0.5]{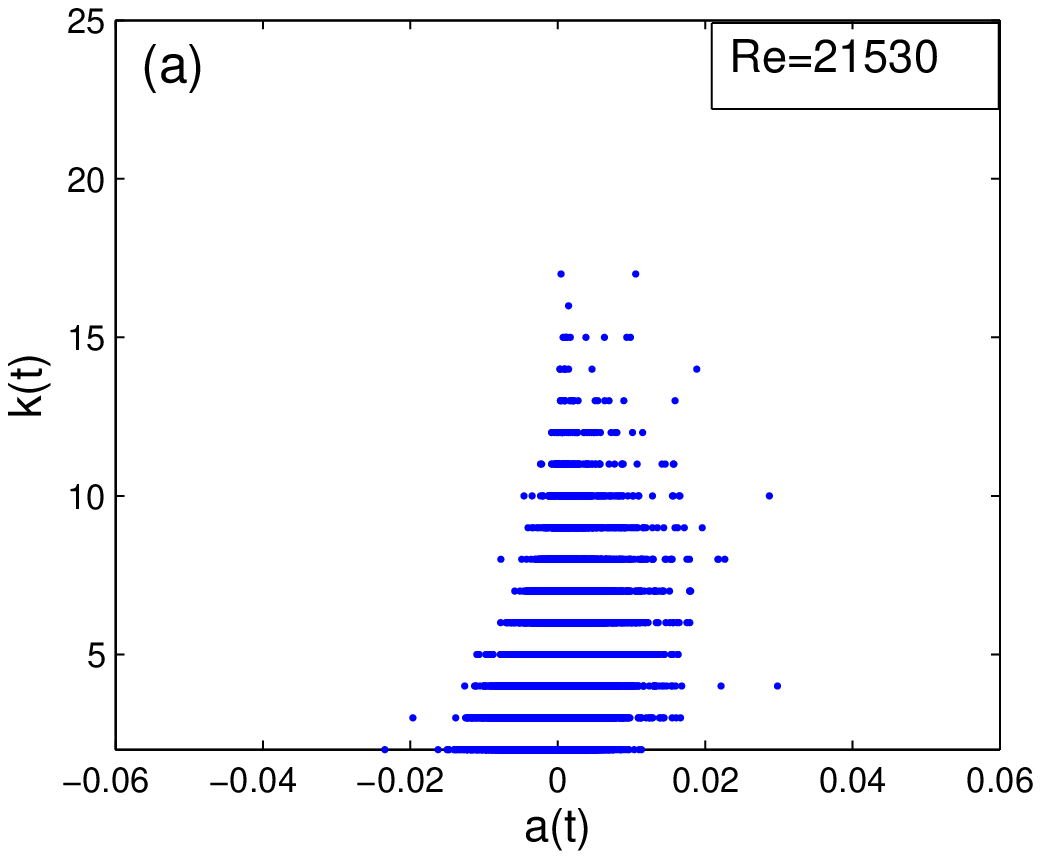}
\includegraphics[scale=0.5]{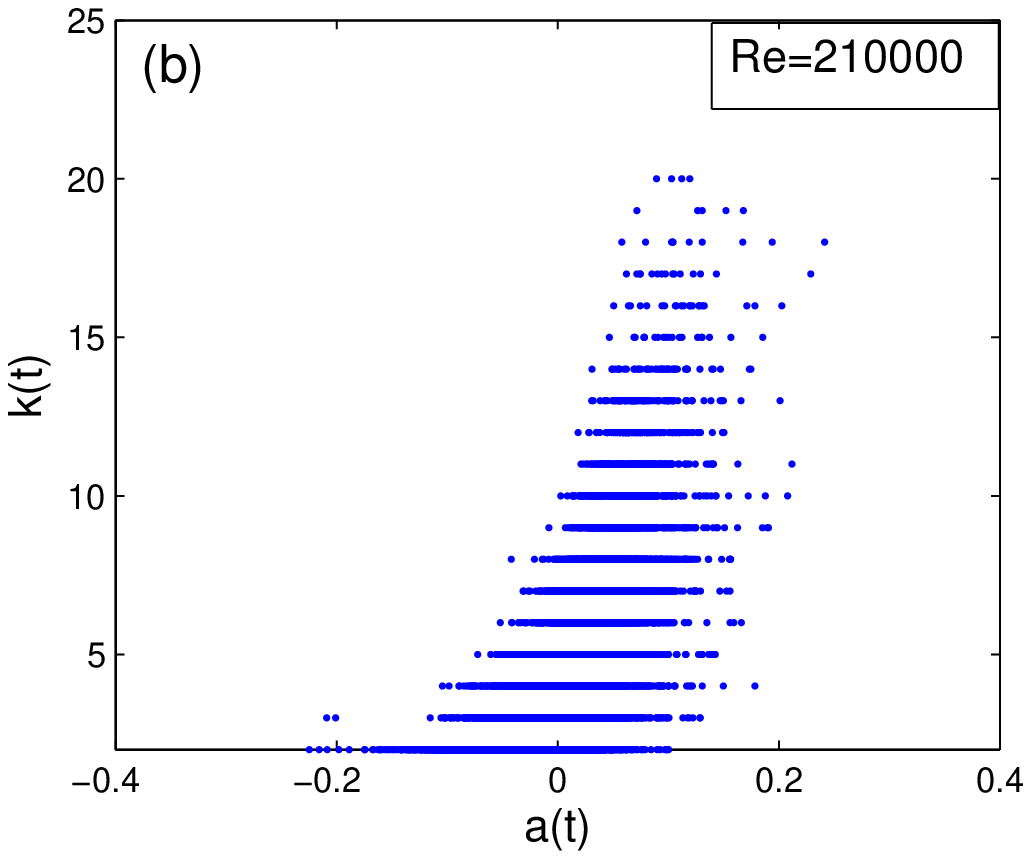}
\caption{The scatter plot of the data point $a(t)$ and its corresponding HVG degree sequence $k(t)$ for two Reynolds numbers (a) $21530$ and (b) $210000$.
The Spearman's correlation coefficients are $0.34$ and $0.74$, respectively.}
\label{k_vs_x}
\end{center}
\end{figure}

We calculate degree assortativity from the adjacency matrix obtained by the HVG algorithm for five different Reynolds numbers, which is shown in Table~\ref{table1}. As it can be seen, the assortativity increases with Reynolds number and the values are positive, which means that there is a tendency for high (low) degree nodes to be connected to other high (low) degree nodes. In other words, the hubs (nodes with highest degree) have better visibility on each other, a phenomenon called \textit{hub attraction} \cite{Song2006}. The presence of hub attraction corresponds to the intermittency or volatility clustering of the original time series.

Further, we also show in Fig~\ref{k_vs_x} the scatter plot between the degree sequence, $k(t)$, and the corresponding data value, $a(t)$ for two Reynolds numbers $21530$ and $210000$. To quantify this dependency, we calculate the Spearman's correlation coefficient \cite{Spearman1904}, $S$, which is a statistical measure of the strength of a \textit{monotonic} relationship between paired data and $-1\le S\le1$. Table~\ref{table1} demonstrates that the Spearman's coefficient increases with Reynolds number and has always positive values. This means that nodes with high degrees correspond to the data points with high values, on average. Due to the obtained values of the degree assortativity (see Table~\ref{table1}), we can conclude that by increasing Reynolds numbers, high degree nodes as well as high-value data points in the original series have a better visibility on each other.

\subsection{Auto-correlation of the degree sequence}
Since the HVG algorithm preserves the time order of data points in the original time series, within the nodes' indices of the graph, we also search
the possible correlation information in the degree sequence, $k(t)$ (see Fig.~\ref{x_t}). Thus, we calculate the lag-one autocorrelation coefficient
of the degree sequence, $\rho_A$, which is defined as follows:
\begin{equation}
\rho_{A}=\frac{\left\langle k(t+1)k(t) \right\rangle -\left\langle k \right\rangle^2}{\sigma_k^2}
\label{auto_corr}
\end{equation}

In Table~\ref{table1}, we present the calculated values of Eq.~\ref{auto_corr} for the five Reynolds numbers. We see that the correlation coefficient of the degree sequence increases by increasing Reynolds number, and their magnitudes are negative. This means that, nodes with high degree are more likely to be followed by nodes with low degree as Reynolds number increases, which is also in agreement with our findings.

\section{Conclusions}
In this article, the statistical properties of velocity and acceleration in turbulent flows have been investigated using the horizontal visibility graph algorithm. We employed this algorithm to map these time series with various Reynolds numbers onto complex networks. At first, we demonstrated that the universal nature of high Reynolds number velocity time series is inherited in the topology of the horizontal visibility network. On the other hand, we found that the degree distributions of the corresponding HVGs for the acceleration series can be modeled by stretched exponential functions. By generating the surrogate data, we also demonstrated that the HVG degree distributions can capture the linear and nonlinear correlation structures in the original time series.

We have also calculated various topological features of the resulting networks, such as the variance of the degrees, the degree assortativity, the autocorrelation coefficient of the degree sequence, and the Spearman's correlation coefficient between the degree sequence and the original time series values. We found a monotonic behavior of the above-mentioned parameters in dependence on $Re$ for a wide range of the Reynolds numbers. However, for high Reynolds numbers, above $R_{\lambda}\simeq 650$, we observed a crossover in different HVG topological properties, which is in agreement with the transitional behavior observed in the recent experimental studies. Then, by calculating the degree assortativity, we showed that the hubs have better visibility on each other (hub attraction) with increasing the Reynolds number. On the other hand, we obtained very high positive values for the Spearman's correlation coefficient, indicating the strong positive relationships between the node's degree and its corresponding value in the original time series.

Due to the hub attraction phenomenon, very high positive values for the Spearman's coefficient, and also negative obtained values for the lag-one autocorrelation coefficient of the degree sequence, we conclude that by increasing the Reynolds number the stochasticity as well as the intermittency of acceleration time series increases. Such results indicate that the horizontal visibility graphs, as a powerful tool for nonlinear time series analysis, can capture the linear/nonlinear correlation structures inherited in the fully developed turbulent flows with various Reynolds numbers.


{\em Acknowledgments---}%
PM  would like to gratefully acknowledge the Persian Gulf University Research Council for support of this work.\\



\begin{thebibliography}{50}

\bibitem{Frisch1995}
Frisch U 1995 \emph{Turbulence: the Legacy of A. N. Kolmogorov} (Cambridge: Cambridge University Press).

\bibitem{Reynolds1883}
Reynolds O 1881 \emph{Phil. Trans. R. Soc. Lond.} \textbf{174} 935.

\bibitem{Batchelor1949}
Batchelor G K and Townsend A A 1949 \emph{Proc. R. Soc. Lond. A} \textbf{199} 238.

\bibitem{Sreenivasan1997}
Sreenivasan K R and Antonia R A 1997 \emph{Annu. Rev. Fluid Mech.} \textbf{29} 435.

\bibitem{Friedrich2011}
Friedrich R, Peinke J, Sahimi M and Rahimi Tabar M R 2011 \emph{Physics Reports} \textbf{506} 87.

\bibitem{Meneveau1991}
Meneveau C and Sreenivasan K R 1991 \emph{J.Fluid Mech.} \textbf{224} 429.

\bibitem{Grossmann1992}
Grossmann S and Lohse D 1992 \emph{Z. Phys. B} \textbf{89} 11; 1993 \emph{Physica A} \textbf{194} 519.

\bibitem{vedula1999}
Vedula P and Yeung P K 1999 \emph{Phys. Fluids} \textbf{11} 1208.

\bibitem{voth1998}
Voth G A, Satyanarayan K and Bodenschatz E 1998 \emph{Phys. Fluids} \textbf{10} 2268.

\bibitem{porta2001}
Porta A L \etal 2001 \emph{Nature} \textbf{409} 1017.

\bibitem{voth2002}
Voth G A \etal 2002 \emph{J. Fluid Mech.} \textbf{469} 121.

\bibitem{Biferale2004}
Biferale L \etal 2004 \emph{Phys. Rev. Lett.} \textbf{93} 064502.

\bibitem{reynolds2004}
Reynolds A M, Yeo K and Lee C 2004 \emph{Phys. Rev. E} \textbf{70} 017302.

\bibitem{Vaillancourt2000}
Vaillancourt P A and Yau M K 2000 \emph{Phys. Bull. Am. Meteorol. Soc.} \textbf{81} 285.

\bibitem{Weil1992}
Weil J C, Sykes R I and Venkatram A 1992 \emph{J. Appl. Meteorol.} \textbf{31} 1121.

\bibitem{pope1994}
Pope S B 1994 \emph{Annu. Rev. Fluid Mech.} \textbf{26} 23.

\bibitem{pratsinis1996}
Pratsinis S E and Srinivas V 1996 \emph{Powder Technol.} \textbf{88} 267.

\bibitem{Farzaneh}
Shayeganfar F \etal 2010 \emph{Phys. Rev. E} \textbf{81} 026304.

\bibitem{Sawford2003}
Sawford B L \etal 2003 \emph{Phys. Fluids} \textbf{15} 3478.

\bibitem{Hill1995}
Hill R J and Wilczak J M 1995 \emph{J. Fluid Mech.} \textbf{296} 247.

\bibitem{Tabeling1996}
Tabeling P \etal 1996 \emph{Phys. Rev. E} \textbf{53} 1613.

\bibitem{Tabeling2002}
Tabeling P and Willaime H 2002 \emph{Phys. Rev. E} \textbf{65} 066301.

\bibitem{Barabasi2002}
Barab\'asi A-L 2002 \emph{LINKED: The New Science of Networks} (Cambridge, Massachusetts: Perseus Publishing).

\bibitem{Barrat2008}
Barrat A, Barth\'elemy M and Vespignani A 2008 \emph{Dynamical Processes on Complex Networks} (New York: Cambridge University Press).

\bibitem{Satorras2007}
Pastor-Satorras R and Vespignani A 2007 \emph{Evolution and Structure of the Internet} (Cambridge: Cambridge University Press).

\bibitem{Murray1993}
Murray J D 1993 \emph{Mathematical Biology} (Berlin: Springer Verlag).

\bibitem{Dezso2002}
Dezs\"o Z and Barab\'asi A-L 2002 \emph{Phys. Rev. E} \textbf{65} 055103(R).

\bibitem{Montakhab2012}
Montakhab A and Manshour P 2012 \emph{Europhys Lett.} \textbf{99} 5800.

\bibitem{Manshour2014}
Manshour P and Montakhab A 2014 \emph{Commun. Nonlinear Sci. Numer. Simul.} \textbf{19} 2414.

\bibitem{Zhang2006}
Zhang J and Small M 2006 \emph{Phys. Rev. Lett.} \textbf{96} 238701.

\bibitem{Lacasa2008}
Lacasa L \etal 2008 \emph{Proc. Natl. Acad. Sci. USA} \textbf{105} 4972.

\bibitem{Lacasa2009pre}
Luque B, Lacasa L, Ballesteros F and Luque J 2009 \emph{Phys. Rev. E} \textbf{80} 046103.

\bibitem{Shirazi2009}
Shirazi A H \etal 2009 \emph{J. Stat. Mech.} \textbf{2009} 07046.

\bibitem{Telesca2012}
Telesca L and Lovallo M 2012 \emph{Europhys Lett.} \textbf{97} 50002.

\bibitem{marwan2009}
Marwan N \etal 2009 \emph{Phys. Lett. A} \textbf{373} 4246.

\bibitem{Donner2011}
Donner R V \etal 2011 \emph{Int. J. Bifurcation Chaos} \textbf{21} 1019.

\bibitem{Lacasa2009epl}
Lacasa L, Luque B, Luque J and Nu\~no J C 2009 \emph{Europhys Lett.} \textbf{86} 30001.

\bibitem{Lacasa2010}
Lacasa L and Toral R 2010 \emph{Phys. Rev. E} \textbf{82} 036120.

\bibitem{Ravetti2014}
Ravetti M G \etal 2014 \emph{PLoS ONE} \textbf{9}(9) 108004.

\bibitem{Renner2001}
Renner C, Peinke J and Friedrich R 2001 \emph{J. Fluid Mech.} \textbf{433} 383.

\bibitem{Donner2012}
Donner R V and Donges J F 2012 \emph{Acta Geophysica} \textbf{60} 589.

\bibitem{Stretched}
Kohlrausch R 1854 \emph{Annalen der Physik und Chemie} \textbf{91} 179; Williams G and Watts D C 1970 \emph{Trans. Faraday Soc.} \textbf{66} 80.

\bibitem{Bogachev2007}
Bogachev M I, Eichner J F and Bunde A 2007 \emph{Phys. Rev. Lett.} \textbf{99} 240601.

\bibitem{Schreiber1996}
Schreiber T and Schmitz A 1996 \emph{Phys. Rev. Lett.} \textbf{77} 635.

\bibitem{Newman2002}
Newman M E J 2002 \emph{Phys. Rev. Lett.} \textbf{89} 208701.

\bibitem{Song2006}
Song C, Havlin S and Makse H A 2006 \emph{Nature Phys.} \textbf{2} 275.

\bibitem{Spearman1904}
Spearman C 1904 \emph{Amer. J. Psychol.} \textbf{15} 72.

\end{thebibliography}
\end{document}